\documentclass[preprint,12pt]{elsarticle}

\usepackage{graphicx}
\usepackage{xspace}
\usepackage{graphicx}
\usepackage{a4wide}

\usepackage{amssymb,amsmath}
\usepackage{amsthm}

\usepackage{lineno}

\journal{Astroparticle Physics}

\newcommand{\keyw}[1]{{\bf #1}}
\newcommand{\hlch}{\bf}
\newcommand{\hlcg}{\bf}

\begin{document}

\begin{frontmatter}



\title{The topological second-level trigger \\of the H.E.S.S. phase 2 telescope}


\author[1]{Y.~Moudden }
\ead{yassir.moudden@cea.fr}
\author[2,3]{A.~Barnacka}
\ead{anna.barnacka@cea.fr}
\author[2]{J.-F.~Glicenstein}
\ead{glicens@cea.fr}
\author[1]{P.~Venault}
\author[1]{D.~Calvet}
\author[2]{M.~Vivier} 
\author[4]{G.~Fontaine} 
\address[1]{{ DSM~/~IRFU~/~SEDI, CEA~/~Saclay, F-91191 Gif-sur-Yvette, France}}
\address[2]{{ DSM~/~IRFU~/~SPP, CEA~/~Saclay, F-91191 Gif-sur-Yvette, France}}
\address[3]{{Nicolaus Copernicus Astronomical Center, Warszawa, Poland}}
\address[4]{{Laboratoire Leprince-Ringuet, \'Ecole Polytechnique, CNRS / IN2P3, F-91128, Palaiseau, France}}

\begin{abstract}
H.E.S.S is an array of atmospheric Cherenkov telescopes dedicated to GeV-TeV
$\gamma$-ray astronomy. The original array has been in
operation since the beginning of 2004. It is composed of four 12-meter 
diameter telescopes. The installation 
of a fifth 28-meter diameter telescope is being completed. This telescope will
operate both in stereoscopic mode and in monoscopic mode \textit{i.e.} without a 
coincident detection on the smaller telescopes. 
A second-level trigger system is needed to supress spurious triggers of
 the 28-meter telescope when operated in monoscopic mode.
This paper gives the motivation and principle of the second-level trigger. 
The principle of operation is illustrated by an example 
algorithm. The hardware implementation of the second level trigger system of 
H.E.S.S. phase 2 is described and its expected performances are then evaluated.
\end{abstract}

%\begin{keyword}
%%Gamma-ray astronomy, imaging atmospheric Cherenkov telescope, trigger, FPGA, real-time, HESS


%\end{keyword}

\end{frontmatter}


%------------------------------------------------------------------------------------------------
\section{Introduction}
\label{sec:intro}
%------------------------------------------------------------------------------------------------
The H.E.S.S. (High Energy Stereoscopic System) instrument is an array of four imaging atmospheric Cherenkov telescopes working in stereoscopic mode. It is located in Namibia, in the Khomas Highland and has started its
operations in 2004. Each of the present ``small'' Cherenkov telescopes (SCT) has a 12-meter diameter mirror and is equipped with a 960-pixel (photomultiplier) camera at its focal plane. It detects photons in the 100 GeV-50 TeV
energy range. 
{\hlch In addition to photons showers, the combinatorial background from diffuse sky photons and charged cosmic showers can trigger the telescopes.
} 
Stereoscopy allows one to achieve a large rejection of the single muon triggers. These single muons come from very distant hadronic showers and dominate the single-telescope {\hlch particle} triggers
\cite{Funk2005}. {\hlcg The single muon trigger rate is discussed in section \ref{sect:part_trig_rates}.} 
The H.E.S.S. collaboration has started to build a fifth, 28-meter-diameter large Cherenkov telescope (LCT).
The LCT will be equipped with a 2048-pixel camera at its focal plane. The LCT will be sensitive to
photons down to 10~GeV.
In normal operation, the SCTs are
triggered only in case of a coincidence with another telescope (LCT or SCT).
However, the energy threshold of SCTs is too high to efficiently detect low energy ($\le 50$ GeV) $\gamma$ rays.
To increase its acceptance at low photon energies, the HESS instrument will have to accept standalone LCT triggers. 
{\hlch Assuming similar first level trigger conditions on the SCTs and the LCT, 
these standalone LCT triggers would have a rate which is typically a factor of five larger than H.E.S.S stereoscopic triggers.
As shown later in section \ref{sect:part_trig_rates}, these triggers are mostly background triggers.}
The H.E.S.S. collaboration has decided to build a second level (L2) trigger board in the camera of the LCT to improve the 
rejection of accidental night-sky background triggers and single muon triggers. 
The L2 trigger board is programmable, which gives flexibility 
in the choice of the trigger algorithms. For instance,
low energy selection algorithms similar to the trigger used by the MAGIC
collaboration to detect the pulsed emission from the Crab pulsar \cite{MAGICCrab} 
can be implemented. These low energy selection algorithms allow to lower the energy threshold on the LCT. 
Alternatively, at constant energy threshold, the gain in bandwidth 
obtained by rejecting the background events can be used to transfer timing 
informations on fired pixels to the acquisition farm, 
in addition to the total charge. The timing information may be useful 
for analyzing {\hlcg single telescope} events, as has also been 
shown by the MAGIC collaboration
\cite{MAGICtiming}. 

Topological triggers have
been previously used on other Cherenkov instruments. For example, the MAGIC collaboration \cite{Bastieri2001}
{\hlch uses a N-next-neighbor logic in its first level trigger and}  
has designed a second-level trigger which can perform a rough event analysis and {\hlcg can apply} topological cuts to the images. 

In the first part of this paper, the various contributions to the LCT instrument trigger rate are listed and evaluated. The next section is devoted to the L2 concept and an example L2
trigger algorithm is given. The actual L2 trigger board is described in section \ref{sect:hardware}. Finally, the on-board 
implementation of the L2 algorithm is discussed in section \ref{sect:firmware}.
%
%------------------------------------------------------------------------------------------------
\section{Level 1 trigger and trigger rates }
\label{sec:l1trig}
%------------------------------------------------------------------------------------------------
%
%
%------------------------------------------------------------------------------------------------
\subsection{Level 1 trigger and contributions to the trigger rate}\label{sect:l1trig_general}
The triggering of the H.E.S.S. phase 1 (HESS-1) instrument has been described in details in reference \cite{Funk2005}. It operates in a two-step process. The first step (hereafter called ``L1 trigger'') is a local
camera trigger. It is a multiplicity trigger in overlapping sectors of 64 pixels.  A camera trigger occurs if the signals in M pixels within a
sector (multiplicity threshold ) exceed a threshold of N photoelectrons (pixel threshold ). The effective time window for pixel coincidence is 1.3 ns. 
The second step is the so-called ``Central Trigger''. The Central Trigger system looks for coincidences of telescope triggers inside a 40 ns time window (``stereoscopic'' events). 
The HESS-1 array is operated in stereoscopic mode. A coincidence of at least 2 telescopes is required
in the Central Trigger time window.

Data acquisition from the large telescope in the phase 2 of H.E.S.S. (HESS-2) will be triggered in three steps.
%The triggering of the phase 2 of H.E.S.S. (HESS-2) will operate in three steps. 
The first step is a camera-level trigger similar to the L1 trigger of HESS-1. 
The camera of the LCT has 96 overlapping trigger sectors.
In addition to time coincidences between SCT 
L1 triggers, the Central Trigger System will check for time coincidences of LCT and SCT triggers. The result of the latter 
coincidence test (monoscopic or stereoscopic event) will be sent back to the LCT
trigger management. As in HESS-1, stereoscopic events will always be accepted. 
In a third step, the LCT monoscopic events are accepted or rejected depending on the result of the L2 trigger system computations.

The largest contributions to the trigger rate of single telescopes in $\gamma$-ray astronomy are background events. 
{\hlch An important source of background events comes from the coincidental quasi-simultaneous firing of pixels 
by diffuse photons, the so-called Night Sky Background (NSB).} The NSB originates
in diffuse sources, such as the zodiacal light or unresolved light from the galactic plane, and light from bright stars. 
The NSB has been measured at the H.E.S.S. site and NSB photoelectron rates were derived for the SCT \cite{Preuss2002}. 
The expected NSB photoelectron rate is $100\pm13$ MHz per pixel at zenith in extragalactic fields. In galactic fields, the single pixel rate is higher and reaches 200-300 MHz
per pixel. The LCT has a larger collection area (596 m$^{2}$ compared to 108 m$^{2}$), but more pixels (2048 instead of 960) and a smaller angular acceptance (3.\ 10$^{-3}$~sr instead of 6.\ 10$^{-3}$~sr). The expected NSB rate per pixel of the LCT is thus expected to be larger by a factor 1.3.

The other source of background is cosmic-ray showers. These showers are induced by interactions of cosmic hadrons (proton, helium) or electrons/positrons in the atmosphere. The typical proton flux above 3 GeV is 600 m$^{-2}$ sr$^{-1}$s$^{-1}$.
Isolated muons from distant hadron showers also trigger single Cherenkov telescopes. These muon triggers dominate the single telescope particle triggers \cite{Funk2005} and are easily rejected by stereoscopic triggers.
%
%------------------------------------------------------------------------------------------------
\subsection{Particle trigger rates}\label{sect:part_trig_rates}
The outputs from the electronics channels of H.E.S.S.  were simulated with realistic photomultiplier signal shapes and 
electronics readout \cite{Guythesis}. The NSB L1 trigger
rate was simulated by adding random photoelectrons to every readout channel. 
Trigger rates caused by NSB single pixel photo-electron rates of 100, 200 and 300 MHz have been calculated. 
The corresponding LCT L1 trigger rates are given in  table \ref{tab:nsb}
for several L1 trigger conditions. Depending on conditions, the estimated rates range from several MHz to less than a few tens of Hz. 
Since the dead time per event of the LCT acquisition is of the order
of a few microseconds, the acquisition rate should be no more than $\sim$ 100 kHz. Table \ref{tab:nsb} shows that some 
L1 trigger conditions (e.g. a pixel multiplicity of 3 with a pixel threshold of 3 and a NSB pixel photo-electron rate of 200 and 300 MHz) lead to
unmanageably high trigger rates.

\begin{figure}[htb]
\begin{tabular}{cc}
\begin{minipage}[b]{0.49\linewidth}
%\centering{\includegraphics[bb=5 5 520 390, scale=0.4, clip = true ]{./figures/protonM4_width2} }
\centering{\includegraphics[bb=5 5 520 390, scale=0.4, clip = true ]{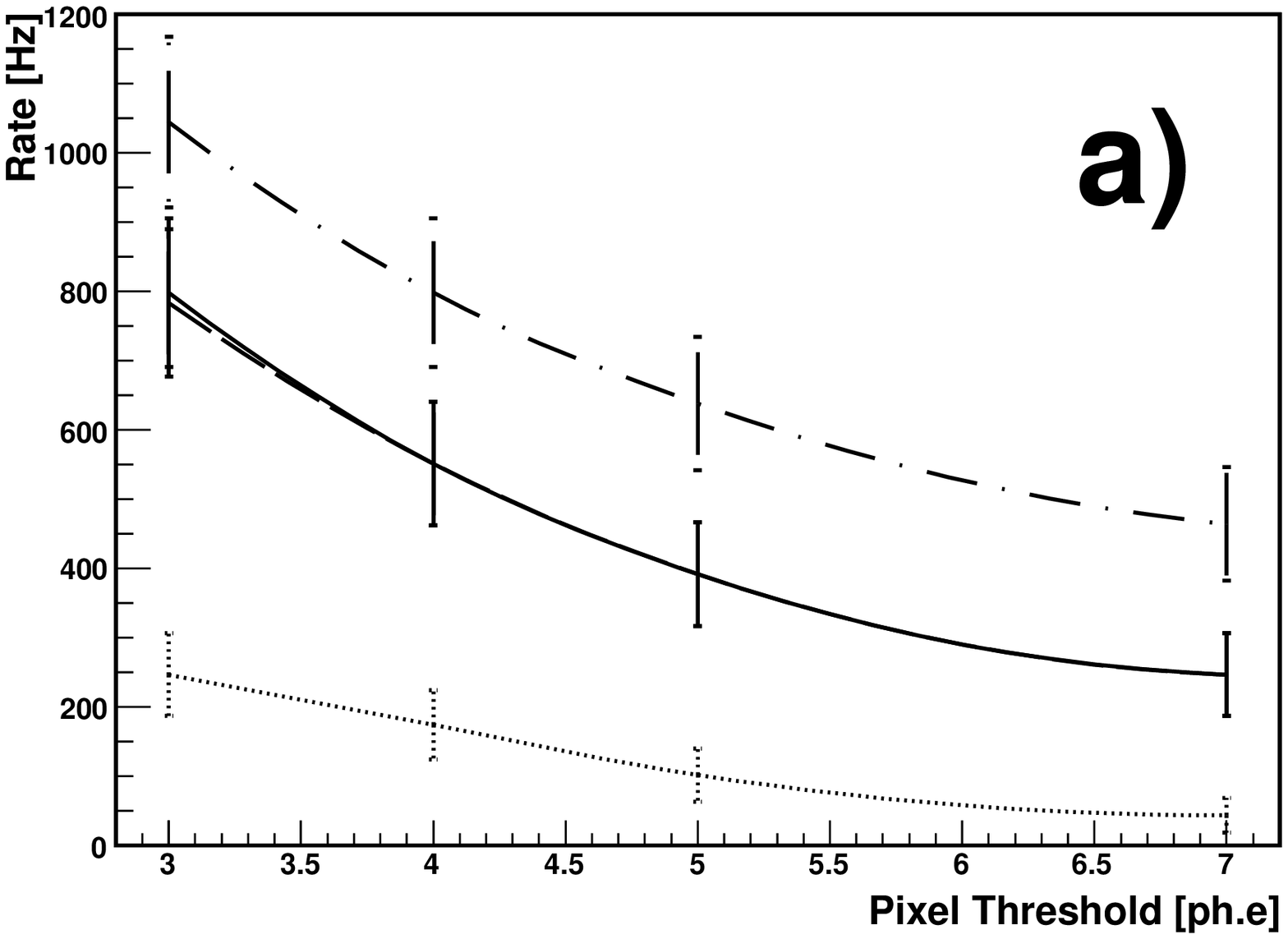} }
\end{minipage}
&
\begin{minipage}[b]{0.49\linewidth}
\centering{\includegraphics[bb=5 5 520 390, scale=0.4, clip = true ]{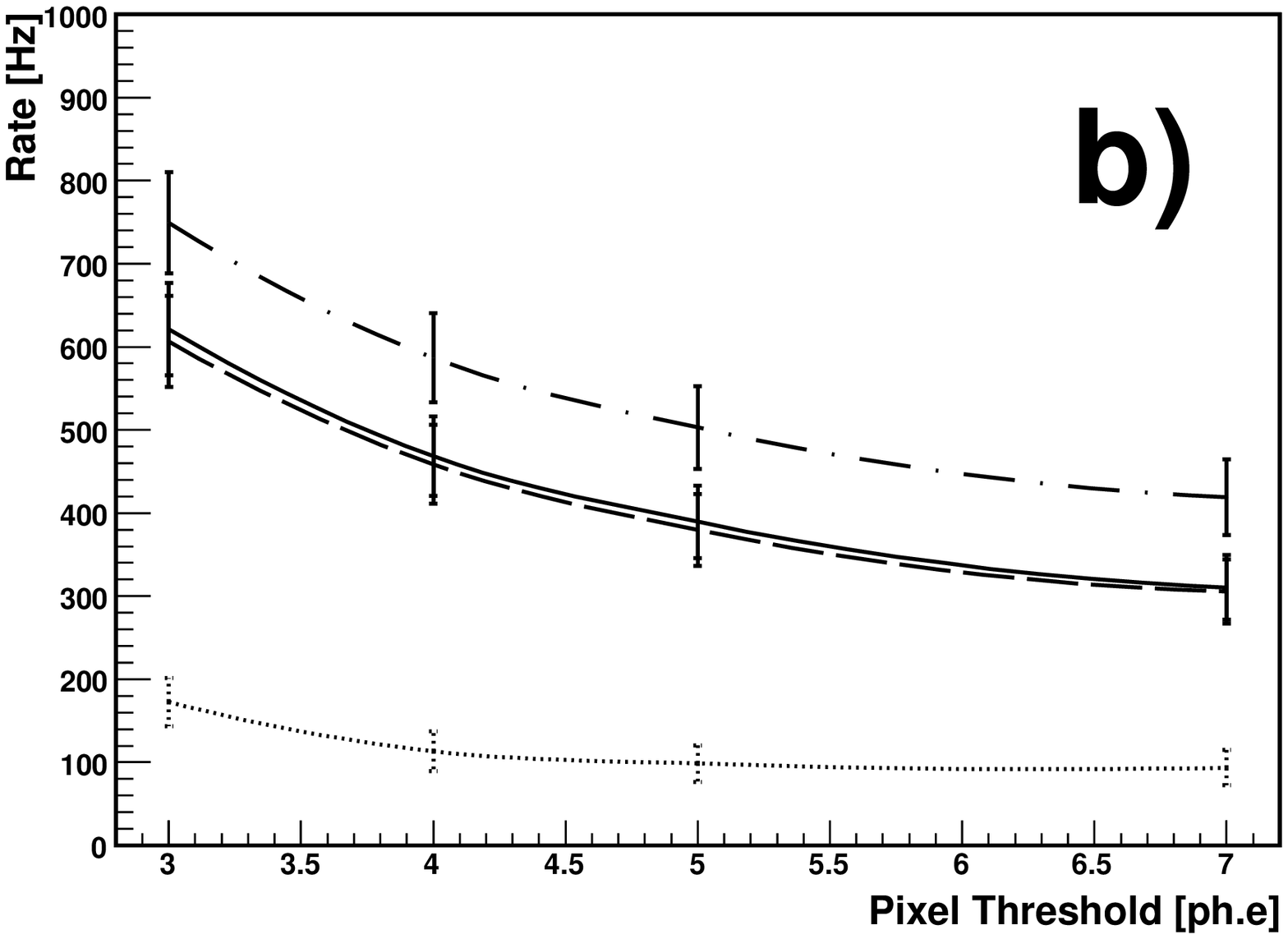} }
%\centering{\includegraphics[bb=5 5 520 390, scale=0.4, clip = true ]{./figures/muonM4_width2} }
\end{minipage}
\\
\end{tabular}
\caption{a) Proton trigger rates versus pixel threshold (in photoelectrons). The pixel multiplicity is 4. b) The part of the trigger rate due to the  isolated muon component of the shower. A L1 pixel multiplicity of 4 and a second level pixel threshold of 7 have been assumed. The dash-dotted line gives the raw L1 rate. The solid line shows the rate of monoscopic events. The dashed line gives the rate of events passing the cleaning/neighboring pixel cut. Finally, the dotted line is the rate of events passing the center of gravity cut. Note that the center of gravity cut reduces the single muon rate by a factor 3. {\hlcg In this figure and figures \ref{fig:electronM4} and \ref{fig:allM4} , the curves drawn are splines connecting the results from simulation.}}
\label{fig:protonM4}
\end{figure}

\begin{figure}[htb]
%\centering{\includegraphics[bb=5 5 520 370, scale=0.6, clip = true ]{./figures/electronM4_width2} }
\centering{\includegraphics[bb=5 5 520 370, scale=0.6, clip = true ]{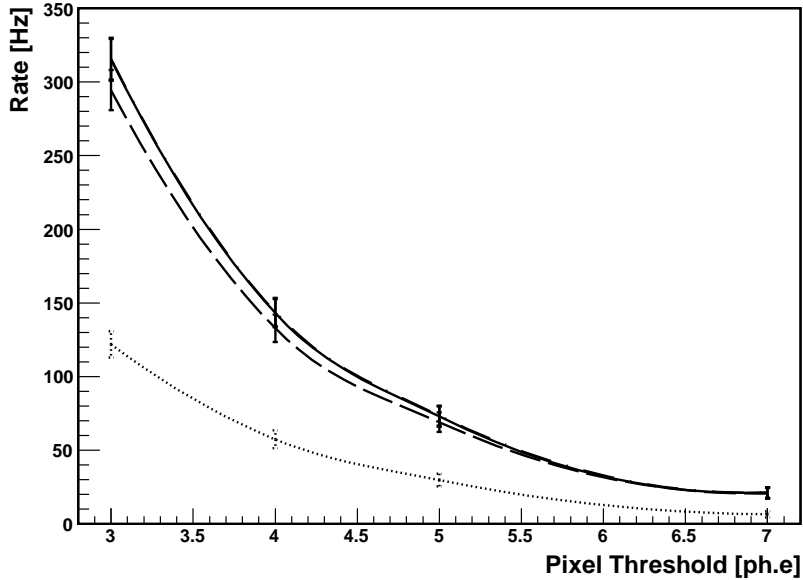} }
\caption{Electron rate as a function of the pixel threshold. A L1 pixel multiplicity of 4 and a second level pixel threshold of 7 have been assumed. The dash-dotted line gives the raw L1 trigger rate. The solid line shows the rate of monoscopic events. These 2 lines are almost superimposed since the electron rate is dominated by low energy events. The dashed line gives the rate of events passing the cleaning/neighboring pixel cut. Finally, the dotted line gives the rate of events passing the center of gravity cut.}
\label{fig:electronM4}
\end{figure}

The {\hlcg proton, muon, photon and electron showers} were simulated with the 
KASKADE \cite{KASKADE} program, using parameterizations given in reference 
{\hlcg \cite{Guythesis}. These parametrisations 
are compared to cosmic ray measurements in reference \cite{Guy2002}.}
The proton trigger rate is shown in figure \ref{fig:protonM4} a) as a function of the pixel threshold in photoelectrons. 
As stated in section \ref{sect:l1trig_general}, a large fraction of the L1 trigger rate is due to
single distant muons. This can be seen by comparing figures \ref{fig:protonM4} a) and b), the latter giving the total L1 rate contributed by isolated muons. Cosmic ray electrons give a Cherenkov signal very similar to the signal of high energy gamma rays. It is thus
not possible to eliminate the electron signal. Further, 
the electron contribution becomes more important at low energy, since cosmic electrons have a very soft 
spectrum (with index $\sim 3.3$). However, the electron background, which is a diffuse source, can be somewhat reduced in 
point source studies. The electron L1 trigger rate is plotted in figure
\ref{fig:electronM4}. An electron rigidity cut-off of 7 GV was assumed \cite{Cortina2001}. The electron trigger rate is a few hundred Hz for
typical L1 trigger conditions.
%
%
%
%\begin{figure}[h!]
%\vspace{2mm}
%\begin{center}
%\hspace{3mm}\psfig{figure=protonM4_width2.eps,width=80mm,angle=0.0}\psfig{figure=muonM4_width2.eps, width=80mm, angle=0.0}
%\caption{a) Proton trigger rate versus pixel threshold (in photoelectrons).
%The pixel multiplicity is 4. b) Part of the trigger rate due to the 
%isolated muon component of the shower.
%A level 1 pixel multiplicity of 4 and a second level pixel threshold of 7 have been assumed. The dash-dotted line gives the raw level 1 rate. The solid line shows the rate of monoscopic events. The dashed
%line gives the rate of events passing the cleaning/neighbouring pixel cut. Finally, the dotted line is the rate of events passing the center of gravity cut. Note that the center of gravity cut reduces the
%single muon rate by a factor of 3.}
%\label{fig:protonM4}
%\end{center}
%\end{figure}
%

\begin{figure}[htb]
%\centering{\includegraphics[bb=5 5 520 285, scale=0.6, clip = true ]{./figures/totalM4_width2} }
\centering{\includegraphics[bb=5 5 520 285, scale=0.6, clip = true ]{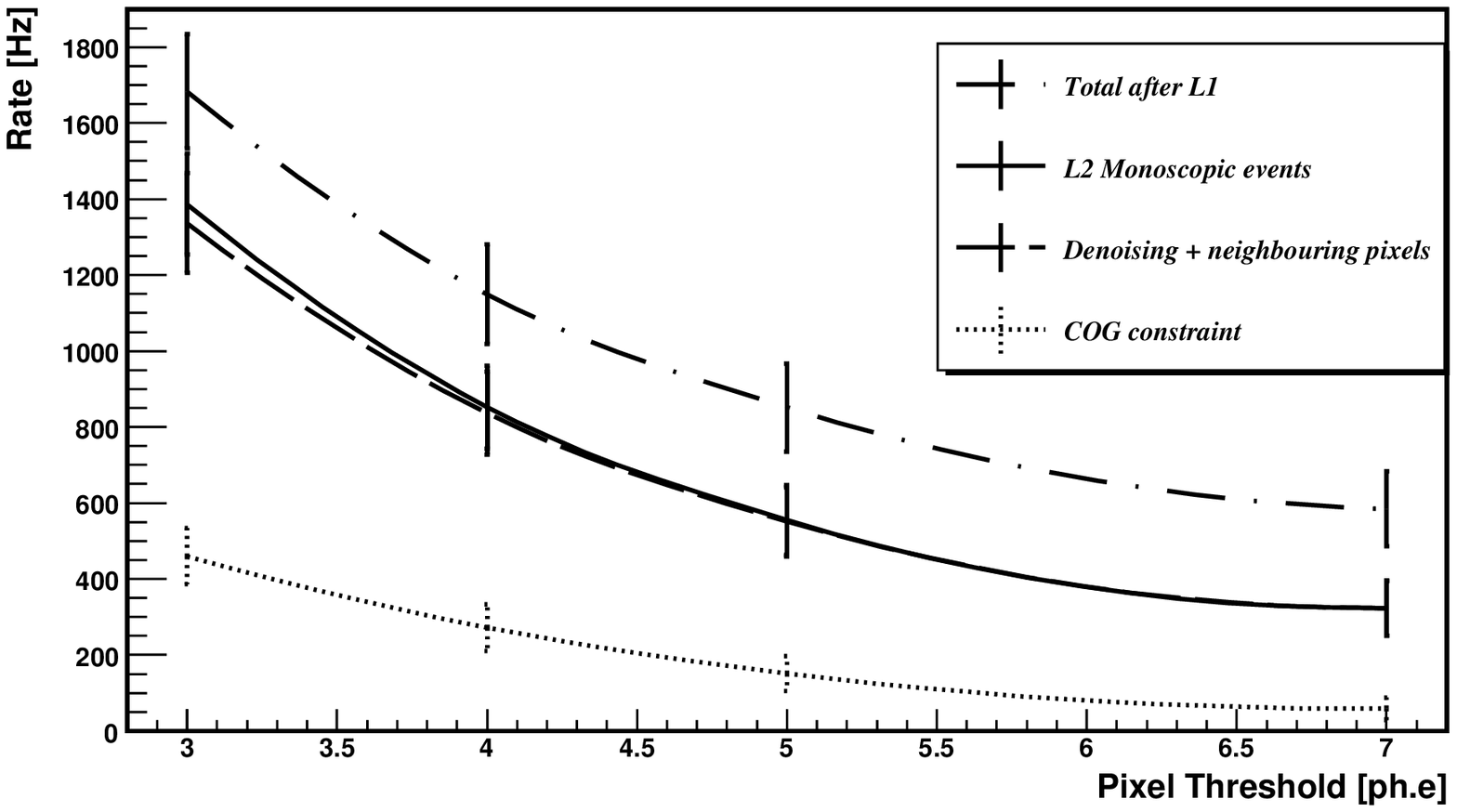} }
\caption{Total hadronic+electron rate as function of the pixel threshold. A L1 pixel multiplicity of 4 and a second level pixel threshold of 7 have been assumed. The dash-dotted line gives the raw L1 rate. The solid line shows the rate of monoscopic events. The dashed line gives the rate of events passing the cleaning/neighboring pixel cut. Finally, the dotted line is the rate of events passing the center of gravity cut.}
\label{fig:allM4}
\end{figure}

The total particle trigger rate is shown in figure \ref{fig:allM4} as a function of the pixel threshold. The particle trigger rate is the sum of the proton, the helium and the electron rate. The helium rate is approximately taken into account by multiplying the proton rate by 0.2 \cite{Guythesis}. The total particle trigger rate is of the order of 1 kHz for typical L1 trigger conditions.
%
%
%\begin{figure}[h]
%\vspace{2mm}
%\begin{center}
%\hspace{3mm}\psfig{figure=electronM4_width2.eps, height=80mm, angle=0.0}
%\caption{Electron rate as a function of the pixel threshold.A level 1 pixel multiplicity of 4 and a second level pixel threshold of 7 have been assumed. The dash-dotted line gives the raw level 1 rate. The
%solid line shows the rate of monoscopic events. These 2 lines are almost superimposed since the electron rate
%is dominated by low energy events.
%The dashed line gives the rate of events passing the cleaning/neighbouring pixel cut. Finally, the dotted line is the rate of events passing the center of gravity cut}
%\label{fig:electronM4}
%\end{center}
%\end{figure}
%
%
%
%\begin{figure}[h]
%\vspace{2mm}
%\begin{center}
%\hspace{3mm}\psfig{figure=totalM4_width2.eps,height=80mm,angle=0.0}
%\caption{Total hadronic+electron rate as function of the pixel threshold.
%A level 1 pixel multiplicity of 4 and a second level pixel threshold of 4 have been assumed. The dash-dotted line gives the raw level 1 rate. The solid line shows the rate of monoscopic events. The dashed
%line gives the rate of events passing the cleaning/neighbouring pixel cut. Finally, the dotted line is the rate of events passing the center of gravity cut}
%\label{fig:allM4}
%\end{center}
%\end{figure}
%
%
%
%
%------------------------------------------------------------------------------------------------
\section{Algorithms for the level 2 trigger system}
\label{section2:algorithms}
%------------------------------------------------------------------------------------------------
%
%------------------------------------------------------------------------------------------------
\subsection{Requirements for the second level trigger}
As explained in section \ref{sect:part_trig_rates}, the input event rate to the L2 trigger system, {\hlch equal to the L1 rate,} is limited 
by the dead-time of the front-end readout board to less than  $\sim 100$ kHz. The output rate of the L2 system 
is limited by the capacity of the Ethernet connection to the acquisition farm.
The maximal output event rate cannot exceed a few (typically 3) kHz. 
%% 1 Gbit/s /2048/2/12 (bits) = 10^{9}/ 4 10^{4} = 2.5 10^{4} 
Table \ref{tab:nsb} shows that the input rate condition sets a
strong constraint on possible L1 conditions. 
For ``admissible'' L1 conditions {\hlch 
(which fulfill the input event rate condition),} 
the NSB rate is strongly reduced 
except possibly when observing special regions of the sky 
(e.g the Galactic Center) with a large NSB. 
For these L1 conditions, the total particle + NSB rate is at the
level of a few kHz. A further reduction of this rate by a factor of 
2 or 3 allows one to fulfill safely 
the L2 output rate condition even in very noisy environments.
%
%------------------------------------------------------------------------------------------------
\subsection{Principle of the second level trigger}
The actual implementation of the L2 system is described in detail in section \ref{sect:hardware}. 
This section describes the ideas underlying its mode of operation. The L2 trigger system uses
the pixel level information (as opposed to the sector level information used in the L1 trigger) to trigger the LCT.
A gray, 2-bit, image of the camera, called ``combined map'', is sent 
to the L2 system whenever the LCT has a L1 trigger. The 2 bit values of 
the combined map correspond to 2 different values of the pixel threshold, 
the L1 pixel threshold $\delta_1$ and a second higher pixel threshold $\delta_2$. 
%The black and white image obtained by taking only the L1 threshold information (resp. the L2 threshold information) are called ${map}_1$ (resp. ${map}_2$).
The black and white maps obtained with these two thresholds are referred to as ${map}_1$ and  ${map}_2$ respectively.
The background rejection is achieved by running event filters, such as the one described in 
subsection \ref{sec:algosoft}, on ${map}_1$, ${map}_2$ or the combined map. Since stereoscopic
events are always accepted, the L2 trigger system operates differently on stereoscopic and monoscopic events. 
When a L1 trigger of the LCT occurs, the Central trigger checks if another
telescope was triggered. If this is the case, then one has a stereoscopic event, which is automatically accepted by the L2 system. 
If on the contrary the event is monoscopic, then it is accepted only if selected by
the L2 trigger event filter.
%
%------------------------------------------------------------------------------------------------
\subsection{Event filters}
\label{sec:algosoft}
The L2 trigger event filters can be divided into two broad classes: clustering/denoising and statistical sums over pixels. 

The clustering/denoising filters aim at removing the NSB contribution to the trigger rate. The denoising filters 
remove all the isolated pixels from ${map}_1$. If the resultant map is empty, then
the event is rejected. There are several possible clustering algorithms. One possibility consists in simply demanding 2 or 3 neighboring pixel hits around a triggered pixel. The effect of the
denoising/clustering on NSB is illustrated by table \ref{tab:nsb}. The clustering algorithm asks for at least 2 triggered pixels neighboring at least one triggered pixel. The NSB trigger rates are seen to decrease by large
factors, in some cases by several orders of magnitude (see e.g the trigger rates for a pixel threshold of 3 photoelectrons). The efficiency of the clustering/denoising event filter allows one to decrease slightly the
L1 trigger threshold and thus to reach a smaller photon energy threshold.
{\hlch For example, the clustering filter allows to use 
the (multiplicity, pixel threshold) = (3,4) L1 trigger condition with
a NSB trigger rate of less than 1 kHz.}   

The proton, electron, and total particle rates are displayed in figures \ref{fig:protonM4}, \ref{fig:electronM4} and \ref{fig:allM4}. These rates are little affected by the clustering cut. The electron rate is
dominated by low energy events, so that most electron events will trigger only the LCT. The second class of filters: statistical sums 
over pixels, can be used to lower the charged cosmic ray background.
{\hlch These filters are run after the clustering filters and the removal of isolated pixels from ${map}_1$ (denoising).}
Several algorithms are currently being investigated. 
In this paper, an example algorithm that can be used to reject a part of the charged particle background is described. 
This algorithm is valid for point
sources or weakly extended sources of photons. 
It is based on two features of the photon signal. First, low energy photons should be detectable only at small impact 
distances of the center of the LCT. Second,
for a given photon energy, there is a correlation between the impact distance of the shower and 
the position of the center of gravity of the image in the camera. 
{\hlch
On the contrary, the center of gravity of single muon events has a flat 
distribution over the camera. Thus a cut on the position of the center of 
gravity of the image in the camera will remove a fraction of the muon events 
proportional to the area that was cut off while
keeping most of the low energy photon events. 
For illustrative purposes, we demand that the center of 
gravity of accepted showers be located at less than 
$1.75^{\circ}/\sqrt{3} = 1^{\circ}$ from the expected position of the 
pointed photon source. 
This cut should remove 2/3 of the single muon showers.
The rejection factor of general hadron showers is expected to be
smaller because showers seen simultaneously by the LCT and one or
several SCTs will be accepted. 
Figure \ref{fig:protonM4} b) shows the rate of single muon
triggers.
The comparison to figure \ref{fig:protonM4} a) shows that single muon
triggers dominate the charged particle trigger rate. 
Roughly 80\% of the muon triggers are
monoscopic events, in agreement
with figure 10 of reference \cite{Funk2005}.
}
Finally, the rate of monoscopic single muon events is reduced by a factor of three when the center of gravity cut is applied. The same reduction applies to the electron background as shown in figure \ref{fig:electronM4}. The
combined effect of the cuts on the charged particle background is summarized in figure \ref{fig:allM4}. As expected, the charged particle rate is reduced by a factor of roughly 3.

\begin{figure}[htb]
%\centering{\includegraphics[bb=5 5 520 370, scale=0.6, clip = true ]{./figures/Efficiency_vs_energy_C6_34} }
\centering{\includegraphics[bb=5 5 520 370, scale=0.6, clip = true ]{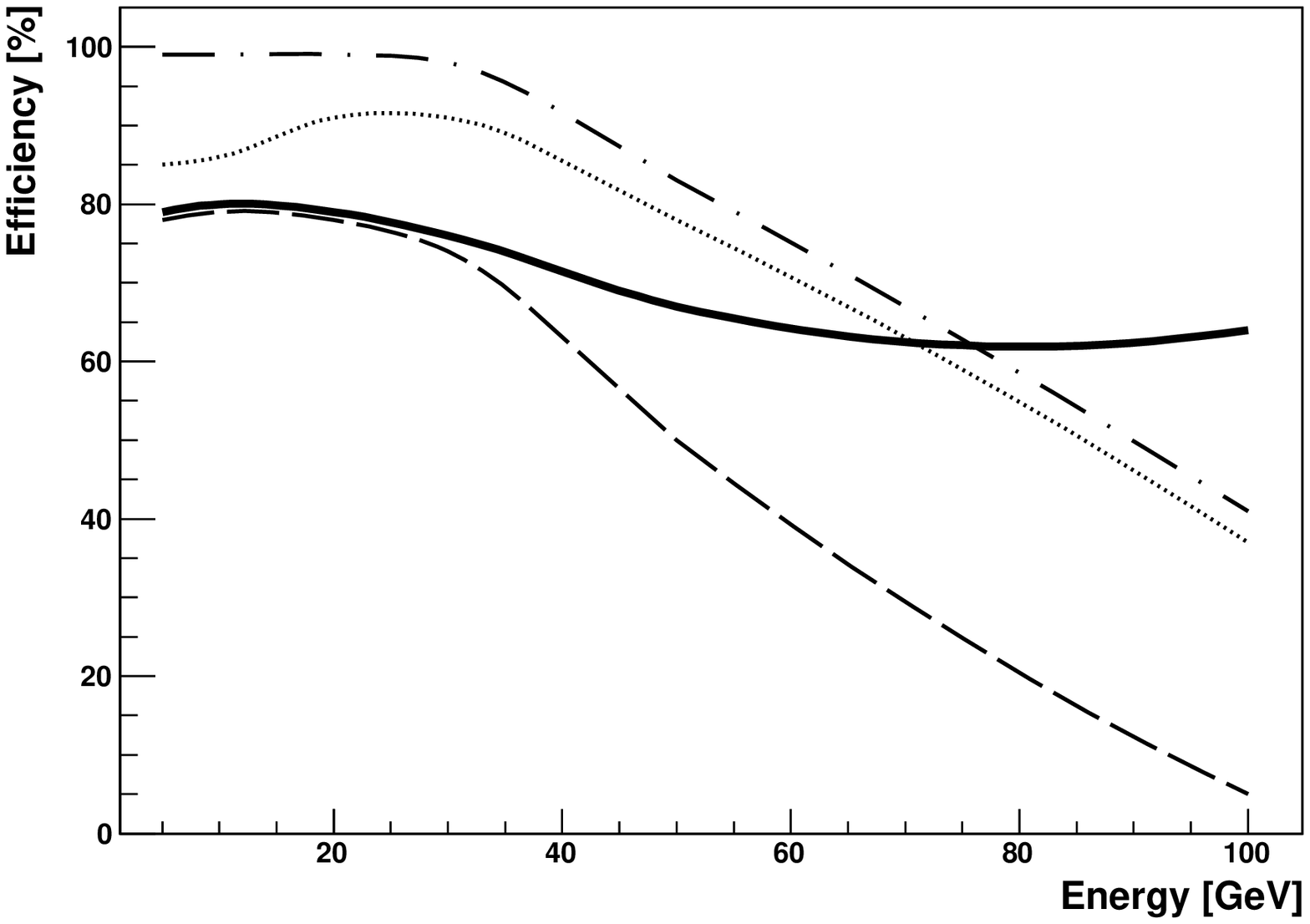} }
\caption{L2 trigger efficiency vs shower energy in the case of filtering by a simple center of gravity algorithm. The efficiency is normalized to the 
L1 photon efficiency. The dot-dashed line shows the fraction 
of monoscopic events.
The dotted line and the dashed line show respectively the effect of the cleaning/nearest neighbor algorithm and the {\hlcg combined effect of the nearest neighbor and center of gravity algorithms.} The L2 efficiency is the sum of the dashed line contribution and of the stereoscopic events.
The fraction of events accepted by the L2 trigger is shown by the solid line.}
\label{fig:efficiecyvsenergy}
\end{figure}

The L2 cuts decrease the photon trigger efficiency. Figure \ref{fig:efficiecyvsenergy} shows the effect of the various L2 cuts 
on the photon efficiency. As the photon energy increases, the fraction of monoscopic events (dot-dashed line) decreases. 
However, the fraction of stereoscopic events, which are automatically accepted by the L2 system, increases. 
The denoising/neighboring pixel event filter (dotted line) removes a 
fraction ($\sim 15\%$) of the 
low energy ( $\le 20$ GeV) photons. After the center of gravity cut 
(dashed line), around $80\%$ of the low
energy photons pass the L2 trigger. The fraction of accepted events 
(solid line) decreases with energy, reaches a minimum of 
roughly $60 \%$ around 75 GeV, 
then increases again because of the increasing fraction of stereoscopic events.

\begin{table}
\begin{center}
\begin{tabular}{cccc}
\hline
(Multiplicity,   & L1 rate      & L1 rate  & L1 rate       \\
Pixel Threshold) & 100 MHz  &   200 MHz               &   300 MHz         \\     
\hline
\hline
(4,3)           & $<$ 63 Hz       & 655 $\pm$ 182 Hz &  \textbf{183 $\pm$ 3.6 kHz} \\
(4,4)           & $<$ 63 Hz       & $<$ 120 Hz      & 142 $\pm$ 51 Hz   \\
(4,5)           & $<$ 63 Hz       & $<$ 120 Hz      & $<$ 160 Hz       \\
(4,5)           & $<$ 63 Hz       & $<$ 120 Hz      & $<$ 162 Hz       \\
(3,3)           & 803 $\pm$ 80 Hz  & \textbf{125 $\pm$ 2.3 kHz}& \textbf{7 $\pm$ 0.18 MHz}  \\
(3,4)           & 84 $\pm$ 40 Hz   & 1 $\pm$ 0.2 kHz  & 16 $\pm$ 1 kHz    \\
(3,5)           & 21 $\pm$ 20 Hz   & 63 $\pm$ 37 Hz   & 1 $\pm$ 0.3 kHz   \\
(3,7)           & $<$ 63 Hz       &$<$ 120 Hz       & 320 $\pm$ 156 Hz  \\
\hline
\hline
\end{tabular} 
\end{center}
%\vspace{0.1cm}
\begin{center}
\begin{tabular}{cccc}
\hline
(Multiplicity,   & clustering& clustering    & clustering  \\
Pixel Threshold) & 100 MHz              &  200MHz                 &   300 MHz\\     
\hline
\hline
(4,3)           & $<$ 63 Hz    & 230 $\pm$ 112 Hz  & \textbf{171 $\pm$ 3.5 kHz} \\
(4,4)           & $<$ 63 Hz    & $<$ 120 Hz       & $<$ 160 Hz  \\
(4,5)           & $<$ 63 Hz    & $<$ 120 Hz       & $<$ 160 Hz  \\
(4,5)           & $<$ 63 Hz    & $<$ 120 Hz       & $<$ 160 Hz  \\
(3,3)           & $<$ 63 Hz    & \textbf{13 $\pm$ 0.24 kHz} & \textbf{8.7 $\pm$ 0.17 kHz}\\
(3,4)           & $<$ 63 Hz    & $<$ 120 Hz       & 510 $\pm$ 212 Hz\\
(3,5)           & $<$ 63 Hz    & $<$ 120 Hz       & $<$ 160 Hz \\
(3,7)           & $<$ 63 Hz    & $<$ 120 Hz       & $<$ 160 Hz \\
\hline
\hline
\end{tabular}
\end{center}
\caption{Night sky background rates. Upper limits are given at the 95\% C.L.
Upper table: Night Sky Background rates for various trigger conditions and {\hlcg NSB photoelectron rates.} Lower table: effect of denoising and clustering. The clustering condition asks for at least 2 neighbors
around at least one triggered pixel. 
{\hlch The L1 trigger rates which exceed the maximum L1 rate of 100 kHz are shown in boldface.}}
\label{tab:nsb}
\end{table}
%
%
%\begin{figure}[h]
%\vspace{2mm}
%\begin{center}
%\hspace{3mm}\psfig{figure=Efficiency_vs_energy_C6_34.eps,width=120mm,angle=0.0}
%\caption{Level 2 trigger efficiency in the case of a simple center of gravity algorithm. The efficiency is normalized to the level 1 photon efficiency. The solid line shows the fraction of monoscopic events.
%The dot-dashed line and the dotted line show respectively the effect of the cleaning/nearest neighbour algorithm and the center of gravity algorithm. The level 2 efficiency is the sum of
%the dotted line contribution and of the stereoscopic events.}
%\label{fig:efficiecyvsenergy}
%\end{center}
%\end{figure}
%
%
%
%
%
%TRANSITION : 
%
In summary, it is possible to efficiently remove the NSB background with a %cleaning/neighbouring pixel 
denoising / clustering algorithm on ${map}_1$. 
The charged particle background can be handled by other, statistical, criteria such as the center of gravity algorithm mentioned above. 
Next, in section~\ref{sect:hardware}, the hardware designed for the L2 trigger system is described. Then, section~\ref{sect:firmware} reports  on firmware and software co-design for the acceleration of the most intensive processing steps of the algorithm and provides experimental timing results.   
%
%--------------------------------------------------------------------------------------------------------------------------
%--------------------------------------------------------------------------------------------------------------------------
%\section{FPGA implementation of the algorithm}
\section{The Level 2 trigger reconfigurable hardware solution}\label{sect:hardware}
%\section{The Level 2 trigger reconfigurable hardware}
%--------------------------------------------------------------------------------------------------------------------------
%The L2 trigger algorithm described above is a possible algorithm, not necessarily realistic. 
Depending on the objective of a given observation run, a different L2 selection algorithm may be preferred.
The L2 trigger system therefore has to be reconfigurable, within certain limits. For the design of the L2 hardware it was assumed that the algorithm 
described in section \ref{sec:algosoft} is representative of other candidate L2 algorithms.
%
%--------------------------------------------------------------------------------------------------------------------------
%\subsection{Reconfigurable hardware solution}
%-------------------------------------------------------------------------------------------------------------------------- 
%
%--------------------------------------------------------------------------------------------------------------------------
\subsection{An original and cost effective hardware solution}
%--------------------------------------------------------------------------------------------------------------------------
%
{\hlch  State-of-the-art technology at the beginning of the L2 trigger board design was provided 
by Xilinx's Virtex4-FX device : an FPGA (field programmable gate array) with an embedded 32-bit PowerPC (PPC) processor 
running at up to 300~MHz\footnote{http://www.xilinx.com/support/documentation/user\_guides}. 
Designing with this platform provides the system with the required flexibility on both firmware and software levels. The auxiliary processor controller unit (APU) of the PPC allows the processor to externalize the execution of custom instructions to the FPGA fabric,  while still using simple function calls in the software. This is a powerful tool for the acceleration of software as shown in section~\ref{sect:firmware}.}

{\hlch A novel non-standard solution was retained for the L2 trigger hardware, consisting in mounting several commercially available boards as mezzanines and a custom designed carrier board. After some  preliminary studies using Avnet's Virtex-4 FX12 Evaluation Kit\footnote{http://www.silica.com/services/engineering/design-tools/ads-xlx-v4fx-evl12-g.htm}
(EB) and  Avnet's Virtex-4 FX12 Mini-Module\footnote{http://www.files.em.avnet.com/files/177/fx12\_mini\_module\_user\_guide\_1\_1.pdf}
(MM) it appeared that these provided the necessary features needed for the  L2 trigger. These boards are distributed by FPGA manufacturers to encourage designers to develop new designs using their latest technology. Hence these evaluation boards are usually very cheap even though they offer a complete  hardware environment for a wide range of applications. In our case, EB was chosen because of the large number of available user I/O's directly connected to the FPGA. Indeed, the two binary maps, sent to the L2 system by the front end (FE) on a Level 1 event, are most easily processed by the L2 algorithm if the whole data converges on a single FPGA. In practice, 64 LVDS links are needed to transfer the images of the 2048 camera pixels from 256 FE boards, grouped into 64 pairs of drawers (\textit{i.e.} sub-crates) as shown in figure~\ref{fig:neighbors}. 
The $\simeq 120$ single-ended cables and the 30 LVDS pairs available on the two 140 pin connectors (AvBus) on the EB provide the necessary connectivity.
Another key feature of the EB is the Micron 32 megabytes DDR SDRAM which the PPC can address to hold code and data.
The mini-module MM, despite its very small footprint ($30~\textrm{mm}$ by  $65.5~\textrm{mm}$), also packages all the necessary functions needed for an embedded processor system. On this board, the Virtex-4 FPGA is accessible through 76 user I/Os and  is connected to 32M x 16 of DDR memory. Both boards hold a 100 MHz oscillator for clocking purposes. This solution greatly reduces the hardware design efforts and most of the firmware can be developed and tested on the evaluation boards concurrently. However, there are a few drawbacks such as coping with the circuit design and the non-standard mechanical format of the evaluation boards. Also the commercial life span of evaluation  boards may be relatively short.}\\
%--------------------------------------------------------------------------------------------------------------------------
\subsection{Custom cPCI carrier board}
\begin{figure}[h!]
\begin{center}
\includegraphics[width=10cm, clip = true]{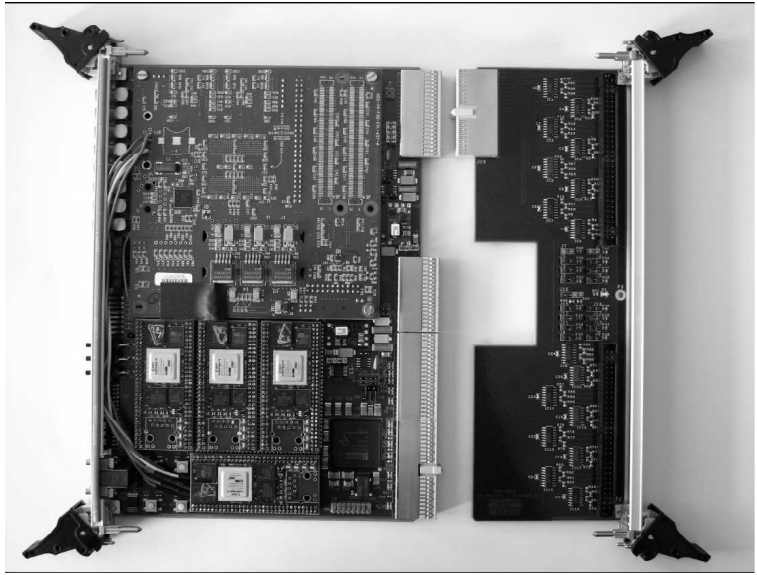}
\caption{%%
View of the final L2 trigger cPCI carrier board equipped with an AVNET EB and 4 minimodules (on the left) and a rear IO board for 
the conversion of 64 LVDS links from FE (on the right).} 
\label{fig:finalhardware}
\end{center}
\end{figure}

%--------------------------------------------------------------------------------------------------------------------------
The mechanical standard of the H.E.S.S. LCT camera is the 6U Compact PCI standard \cite{CPCINORME}.
The designed cPCI  board is shown on the left of figure~\ref{fig:finalhardware}, carrying
four MMs and one EB. 
The  6U rear I/O board, shown on the right of figure~\ref{fig:finalhardware}, is in charge of translating the 64 incoming LVDS pairs to unipolar signals and of forwarding these to the front side of the crate to the \textit{main} L2 trigger board. 
The carrier board 
%supplies the bare essentials. Several regulators 
provides the mezzanines with the different voltage levels needed. 
Communication over the PCI bus is ensured by a PCI bridge located inside a Spartan 3AN FPGA\footnote{http://www.xilinx.com/support/documentation/spartan-3an\_user\_guides.htm}. 
{\hlch The latter also holds the necessary logic to communicate with the FPGAs on the mezzanines for data acquisition, slow control, FPGA configuration and software download by JTAG \textit{via} PCI.}
%
%Indeed, the Virtex-4 FX12s and the external SDRAMs holding the software running on the PPCs are configured by JTAG \textit{via} PCI. 
%
Another important feature provided by Virtex4-FX FPGAs are the special logic blocks for high-speed serial connectivity called I/OSERDES which are available in all IO tiles. In our design with multiple Virtex4-FX devices, this is crucial for FPGA interconnectivity. 
The EB is in charge of receiving the data from the FE, through the backplane. Additional information about the stereoscopic nature of the incoming event reaches the EB through the front panel. 
Fast serial links connect the EB to the MMs so that some data processing tasks can be exported to the other FPGAs available in the L2 trigger system. 
Four slots are reserved for the L2 system in the trigger crate  of the LCT camera and fast serial links are included in the design for interslot data transfers through the backplane. With the current board design, the computing resources of up to 20 Virtex-4FX12 FPGAs are available for the implementation of the L2 trigger algorithm.
%
%-------------------------------------------------------------------------------------------------------------------------------------------
\section{Firmware and Software for the L2 trigger system }\label{sect:firmware}
%--------------------------------------------------------------------------------------------------------------------------------------------
Decisions relative to the design of the L2 trigger board were made based on the good timing results obtained while running an optimized implementation of the default L2 trigger algorithm on the Virtex4-FX evaluation boards.
This insured simple portability of the algorithm onto the final L2 trigger hardware. 
This section reports on the acceleration of the most computationally intensive steps in the L2 algorithm namely the detection of clusters in the data and the computation of Hillas statistics~\cite{hillas}.
Experimental timing results are given using a single L2 trigger board in two different configurations involving either one or five Virtex4-FX FPGAs.
%---------------------------
%---------------------------
\subsection{Deserializing and synchronizing the data}\label{sect:deser}
\begin{figure*}[t!]
\begin{tabular}{cccc}
\begin{minipage}[b]{0.4\linewidth}
\centering{\includegraphics[bb=375 100 830 555, scale=0.43, clip = true]{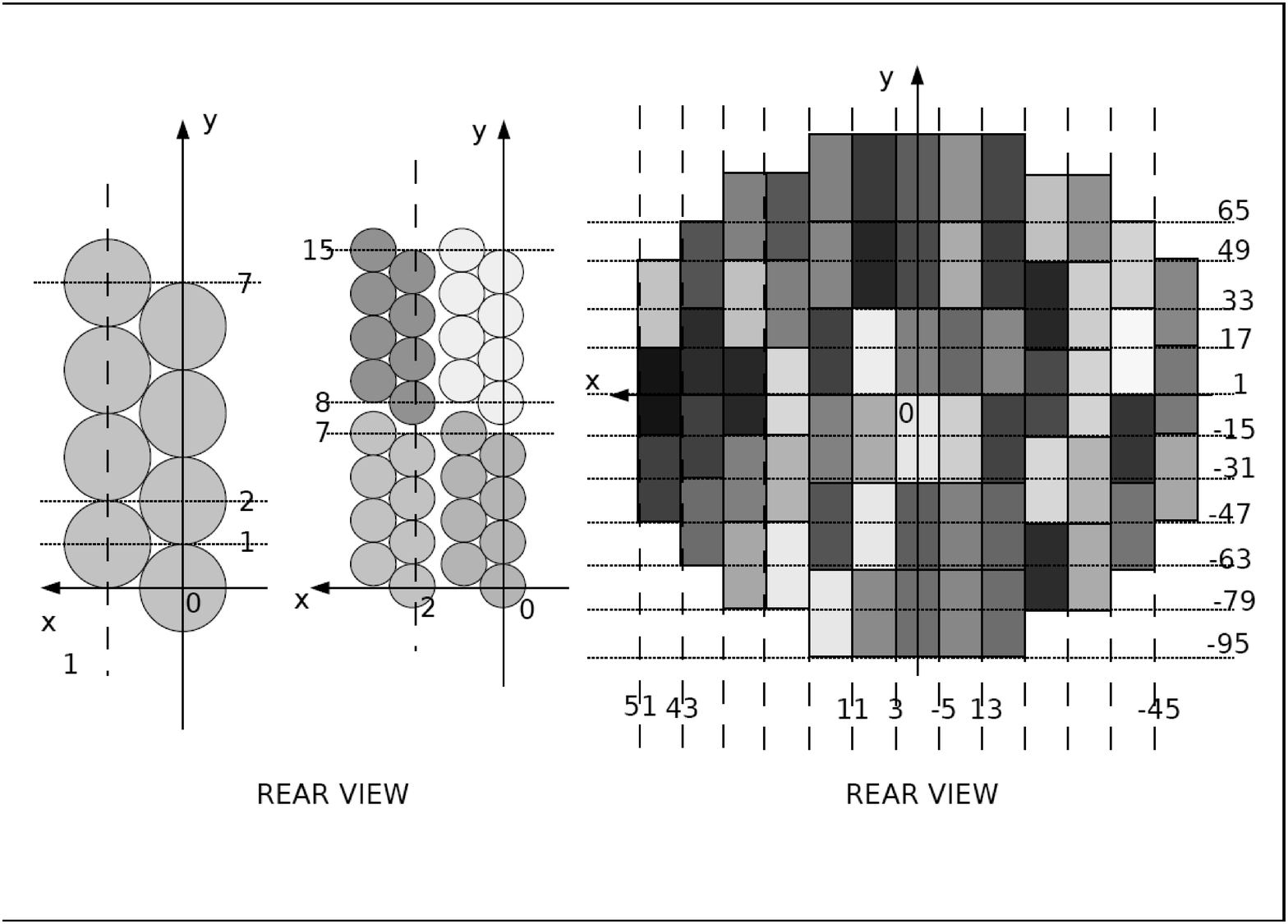}}
\end{minipage}
&
\begin{minipage}[b]{0.15\linewidth}
\centering{\includegraphics[bb=180 100 375 520, scale=0.43, clip = true]{coordinates}}
\end{minipage}
&
\begin{minipage}[b]{0.15\linewidth}
\centering{\includegraphics[bb=5 100 180 520, scale=0.43, clip = true]{coordinates}}
\end{minipage}
&
\begin{minipage}[b]{0.3\linewidth}
\centering{\includegraphics[bb=630 0 880 450, scale=0.43, clip = true]{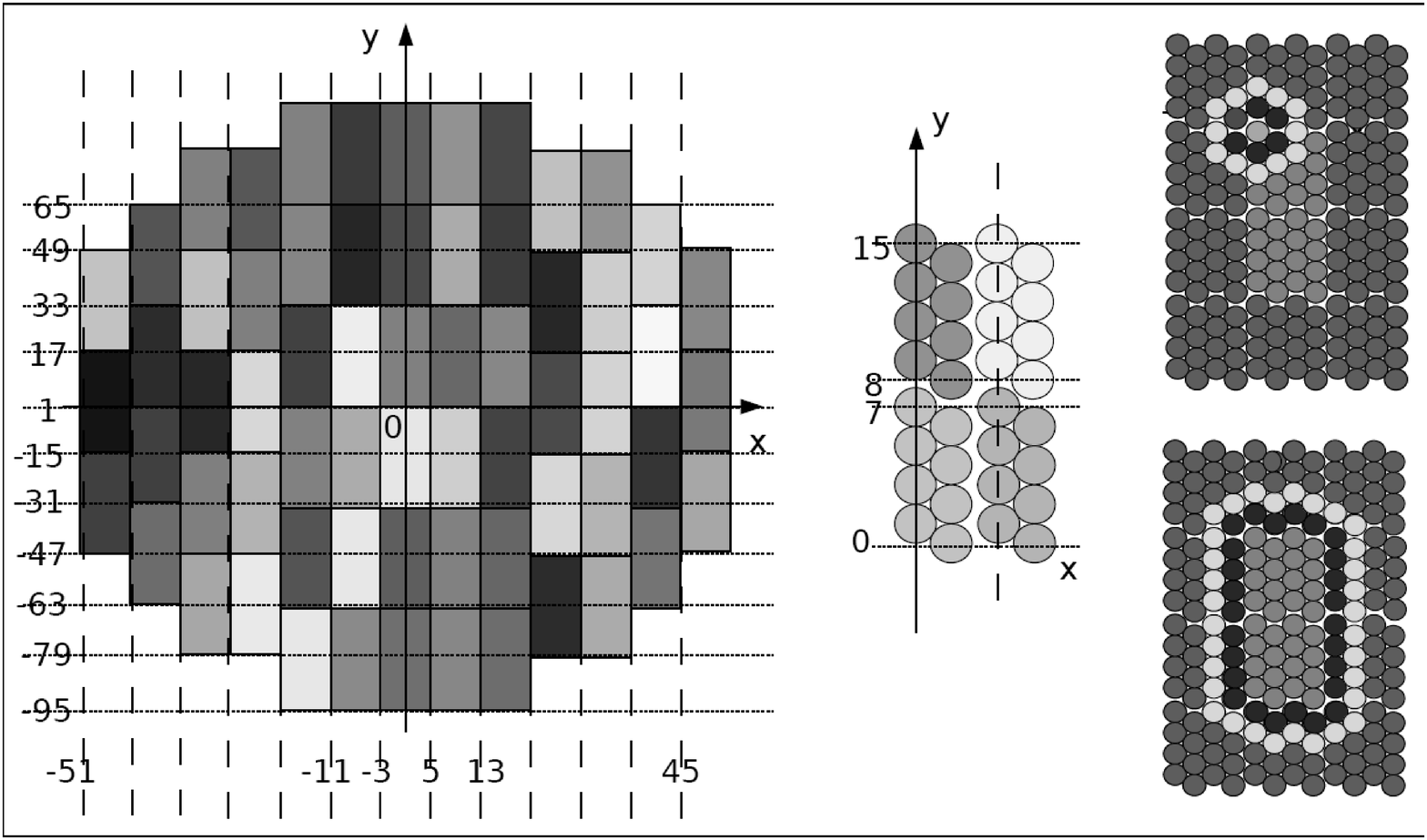}}
\end{minipage}
\\
\end{tabular}
\caption{%%
\textbf{Left~:} Each rectangle represents one of the 64 pairs of drawers (\textit{i.e.} sub-crates) which compose the camera of the LCT. Each pair of drawers contains 4 front end boards and each FE board carries 8 PMs. The 2048 pixels of the camera are on an equilateral triangular grid. However, for simplicity, each PM has integer coordinates in the non orthonormal set of axes shown.  \textbf{Middle~:} Local integer coordinate frames for a simple computation of Hillas parameters from a single FE board, and then from 4 FE boards in the same pair of drawers. \textbf{Right~:} First and second neighbors to the 32 PMs on the 4 FE boards from the same pair of drawers : there are 26 and 32 first and second neighbors to take into account when applying the denoising and clustering filters.}
\label{fig:neighbors}
\end{figure*}

%---------------------------
%---------------------------
The L2 trigger system is in charge of deserializing and  synchronizing the data it receives from the FE and the 
central trigger following a L1 trigger. 
A custom asynchronous serial protocol is used to transfer 4096 data bits from the FE to the L2 trigger system onto 64 LVDS links. Each link is used to transfer 64 bits from 32 pixels on 4 adjacent FE boards as shown in figure~\ref{fig:neighbors}. 
The data are sent as 4 words of 16 bits, each preceded by a start bit and two ID bits. Symbols last 45~ns and a gap of at least 270~ns separates two successive words of 19 bits each so that it takes 4.18~$\mu$s to transfer the data from the FE to the L2 trigger.
%
%The L2 trigger uses adjustable delay lines to synchronize the start bits on the 64 input lines to within the 11.25~ns four-fold oversampling period. 
%
The binary data are then stored in a pipeline as a $64\times64$ bit matrix.  
Concurrently, the central trigger informs the L2 of the stereoscopic nature of each L1 event.
This information is stored in a different FIFO, along with the contents of several registers driven by slow control. The latter are parameters used in the trigger algorithm (\textit{e.g.} target coordinates).  
%
%Dual-port block RAMs were used to implement the above two concurrent data paths. 
%

{\hlch The L2 trigger system as a whole is structured as a pipeline~: it is compelled to provide the local trigger management module with its decisions to accept or reject L1 events in the very order that these L1 events occurred. This is due to the data acquisition FIFOs used to hold the camera data on the FE boards while awaiting the L2 trigger decision. Note that these FIFOs have a capacity of 50 events which sets an upper bound on the latency of the L2 system.}\\
The PPC on the EB is in charge of synchronizing the FE data and the central trigger data by checking the contents of the two BRAMs on its 64 bit wide local bus (PLB).
There are different ways for the PPC to access the BRAMs in the FPGA fabric. However the best timing was obtained using a cacheable BRAM on the PLB. 
If the event to be processed is tagged as \textit{stereo}, the L2 system issues an \textit{accept} signal. {\hlcg Otherwise, the L2 trigger issues either an \textit{accept} or a \textit{reject} decision as an output of the following L2 algorithm~:} 
\begin{tabbing}
\quad \=\quad \=\quad \kill
1. Set to 0 all pixels in $map_1$  that are not in clusters of 3 at least $\rightarrow \widetilde{map}_1$\\
2. \keyw{IF}     $\widetilde {map}_1 = 0$ \keyw{THEN} Reject \keyw{ELSE} \\
\>3. Set to 0 all isolated pixels in $map_1$ $\rightarrow \widehat{map}_1$\\ 
\>4. Compute Hillas statistics of $ \delta_1 \widehat{map}_1  +( \delta_2 -\delta_1 ) map_2 $\\
\>5. Compute distance $\Delta$  from center of gravity $cog$ to target $(x_c, y_c)$\\
\>6. \keyw{IF}     $\Delta \ge  \tau_{\textrm{cog}}$  \keyw{THEN} Reject  \keyw{ELSE}  \textit{ Accept}
\end{tabbing}
where as before ${map}_1$ and ${map}_2$ are the two input binary maps associated with threshold values $\delta_1$ and $\delta_2$,  
 $(x_c, y_c)$ are the pointed target's coordinates on the camera plane and $\tau_{\textrm{cog}}$ is the decision threshold on the nominal distance $\Delta$ between the center of mass of the event maps and the target.

%--------------------------------------------------------------------------------------------------------------------------
\subsection{Hardware acceleration of clustering and denoising }\label{sect:acceler}
%--------------------------------------------------------------------------------------------------------------------------
%
A purely sequential software implementation of the clustering and denoising operators in steps 1 and 3 is clearly time-consuming and sub-optimal. 
We thus turned to a  parallel hardware implementation~: considering the binary nature of ${map}_1$, the non-linear filters involved are easily built using logic AND and OR gates. For instance, denoising is achieved by convolving $map_1$ with a simple \textit{filter} such that~:
\begin{equation}
\widehat{map_1}(i) = map_1(i) \wedge \big(  \bigcup_{j=1..6} map_1(i_j)  \big)
\end{equation}
where $i_j$ is used to index the 6 nearest neighbors of pixel $i$. A wider \textit{filter} with a support extending to second neighbors is used to detect clusters of at least three pixels as shown in figure~\ref{fig:neighbors}.
In order to process 32 pixels from 4 adjacent FE boards in parallel, one needs the values of 58 first and second neighbor pixels from 12 neighboring FE boards as shown in figure~\ref{fig:neighbors}. On the edges of the camera, one or more bytes are set to zero. \\
We note here that filtering ${map}_1$ is a local operation. For this reason, the L2 data pipeline reorganizes the input $64\times64$ binary data matrix to provide local access to ${map}_1$ and ${map}_2$.
This way, each 32-bit word read by the PPC gives the binary values of 32 pixels from one of the 64 groups of 4 adjacent FE boards of the LCT camera and each byte maps to one FE board, as shown in figure~\ref{fig:neighbors}.\\
In practice, there are different possibilities for connecting the proposed logic filter to the PPC. The best timing results were obtained using the APU~\cite{apu}~: experimentally, denoising and 3 pixel cluster detection take $\approx 7500$ PPC clock cycles with using a custom peripheral on the PLB while it takes only $\approx 6200$ cycles using the APU.
%
%--------------------------------------------------------------------------------------------------------------------------
\subsection{Fast computation of $1^\textrm{st}$ and $2^\textrm{nd}$ order moments }
%--------------------------------------------------------------------------------------------------------------------------
%
Computing the first and second order moments of the denoised combined map is a 
common  preliminary for the estimation of Hillas statistics and other parameters of interest~\cite{hillas}. Let us define~:
\begin{eqnarray}
m_x =  \sum_i m_i x_i  \quad & \quad m_y =  \sum_i m_i y_i\\  
m_{xx} =  \sum_i m_i x_i^2 \quad & \quad m_{yy} = \sum_i m_i y_i^2 \\
m_{xy} = \sum_i m_i x_i y_i \quad & \quad m = \sum_i m_i
 \end{eqnarray}
where $i$ indexes the 2048 pixels in the processed data maps, and $m_i$ is the weight assigned to pixel $i$. 
Again, a fully sequential algorithm to compute these quantities is excluded. 
Fortunately, a faster computation is possible thanks to the structure of the $64\times64$ binary data matrix described in the previous section and to the linearity of the above quantities with respect to the pixel weights. Hence the binary maps  $\widehat{map}_1$ and  ${map}_2$ can be processed separately and the sums are profitably rearranged for an efficient hierarchical computation of the moments. First a byte-addressable \textit{look-up table} (LUT)  is used to compute the desired statistics on each FE board in a local coordinate frame attached to the 8 pixels on a single FE board as shown in figure~\ref{fig:neighbors}. The LUT outputs are then combined locally to compute these statistics for the four FE boards in each of the 64 pairs of drawers.  
This local summation requires the LUT outputs to be properly \textit{translated} depending on the position of a given FE board in the current pair of drawers. As shown in figure~\ref{fig:neighbors}, a different coordinate frame is used when handling the 32 pixels in a pair of drawers.
Finally, summation over the 64 pairs of drawers requires an additional transformation of these partial statistics to account for the translation and scaling of the local frame required to move a pair of drawer to its correct position within the global coordinate frame of the camera. 
In the end, the contributions of the two binary maps are linearly combined with weights $\delta_1$ and $ \delta_2 -\delta_1$ providing final 32 bit integer statistics $m_{yy}$, $m_{xy}$, $m_{xx}$, $m_y$, $m_x$ and $m$ for the combined map. 
With this fast implementation, the PPC computes the first and second order moments of the input data in a maximum of 18000 clock cycles. 
For a faster execution time, given that these statistics will most often be estimated for low energy events when only very few pixels are high in $\widehat{map}_1$ and even less in  ${map}_2$, it is worth checking if a byte is zero prior to computing its contribution to the statistics. As a result the computation time will vary almost linearly with the number of \textit{active} bytes in the data.\\ 

The algorithm proposed in section~\ref{sec:algosoft} uses only the first-order statistics to compute the nominal distance obviously in finite precision~:
\begin{equation}
\Delta = \sqrt{  (\frac{ m_y}{ m} - y_c ) ^2 +  3 \times ( \frac{m_x}{ m } - x_c ) ^2}
\end{equation}
where the factor $3$  is due to the equilateral triangular grid and the accompanying $\sqrt 3 $ left out in the moment computation for simplicity. The specified precision on the target coordinates is $1/32^\textrm{th}$ of the unit length, motivating the precision to which the center of gravity ($cog$) coordinates have to be computed. Finally, thresholding the squared nominal distance avoids the costly computation of a square root.  
%
%
%-----------------------------------------------------------	
\subsection{Experimental timing results }\label{sect:results}
%-----------------------------------------------------------	
%
The accelerated blocks described above (leaving out second order moments)  were readily integrated into a fully functional L2 trigger system running on the specifically designed hardware. 
In practice, two simple architectures were tested for preliminary timing experiments in order to benchmark the performance of the L2 trigger system. 
The first design implements the full default algorithm inside a single FPGA. 
The second design uses the five Virtex4-FX FPGAs available on one L2 trigger board. In this case, the Virtex4 on the EB receives the data from the FE, forwards it to the MMs and reads back their decisions. 
%
%Using the fast serial links IOSERDES, transferring the $140*4$ data bytes for one L1 event takes less than 10~$\mu$s at $533~\textrm{Mbits}/\textrm{s}$.
%
Events are assigned to MMs in a simple \textit{round robin} procedure. %, which still needs to be optimized.    
Each MM is in charge of executing the complete decision algorithm. In this simple design, there is no pipelining of events in the MMs.     
For these experiments, an additional evaluation board is used to emulate the L2 trigger's environment. This testbench is capable 
of generating L1 trigger events periodically at a specified rate, as well as bursts of specified lengths. 

{\hlch Consider first the unrealistic worst case scenario where all events are monoscopic   with all pixels high above the higher threshold $\delta_2$ resulting in the longest processing time.} With this setup  a stable behavior of the first  {\hlch single FPGA} system was observed up to a maximum L1 rate slightly above 10~kHz. The second multi-FPGA system could sustain a maximum mean rate close to 30~kHz.
In fact, the additional communication tasks between the EB and MMs in the multi-FPGA system are in part responsible for the less than four-fold gain. Adding more MMs to the system will only increase the maximum acceptable L1 rate to the point where the slowest concurrent process in the pipeline can handle it, which is $\approx 60$~kHz in the current multi-FPGA design.\\  

\begin{figure}
\begin{center}
\includegraphics[bb=5 5 520 390, scale=0.5, clip = true]{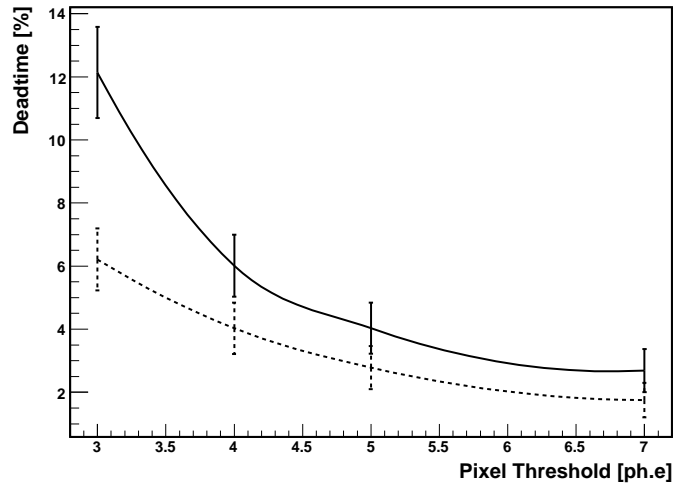}
\caption{%%
Estimated deadtime of the PPC in the single FPGA implementation of the L2 trigger system. The continuous and dashed lines correspond to  L1
pixel mulitiplicities equal to 3 and 4 respectively.}
\label{fig:maxmeanrate}
\end{center}
\end{figure}

The above estimates are clearly highly pessimistic as they are based on unrealistic data.
A more realistic estimation of the PPC occupancy is plotted in figure~\ref{fig:maxmeanrate} for the first 
implementation of the L2 trigger system. This estimate is based on the simulated statistical distribution of input events  
reported in section~\ref{sec:l1trig} and on time measurements of the different elementary steps in the L2 trigger pipeline. 
For typical L1 trigger conditions - multiplicity set to 3 or 4 and pixel threshold between 3 and 7- 
the estimated average processing time is  
$\sim$ 37~$\mu$s which corresponds to a maximum mean L1 rate of 27~kHz.  
Actually, for these trigger conditions, the system occupancy is estimated $\le 20$ \% as shown in figure~\ref{fig:maxmeanrate}.
{\hlch The mean latency of the L2 trigger in its single FPGA implementation is the average time spent by an event inside the L2 pipeline. This includes steps before (\textit{i.e.} data reception and data matrix transposition ) the sequential processing by the PPC and after (\textit{i.e.} transmission of L2 decision). Experimentally, we measured the average L2 latency to be $\approx 38~\mu\textrm{s}$.}
The proposed multi-FPGA system will obviously provide a larger safety margin in terms of PPC occupancy as well as a shorter latency. 
In all cases the definite real L1 rate will have to be determined on site.
%
%
%------------------------------------------------------------------------------------------------
\section{Conclusion}
%------------------------------------------------------------------------------------------------
%
This paper describes the design and implementation of the L2 trigger system for the second phase of the H.E.S.S. experiment.
The L2 trigger will be used to reject night sky background related and isolated muon events and thus reduce the trigger rate. 
The principle of the trigger is to build a 2-bit (``combined'') map of the camera pixels at the time of trigger. The night sky 
background events can then be rejected by demanding clusters of pixels on the combined map. Further rejection of the hadronic background
can be obtained by using quantities such as the center of gravity of the triggered pixels. A possible, illustrative, algorithm for the L2
trigger system has been given in section \ref{section2:algorithms}. This example algorithm shows that the required rejection
of night sky background and isolated muon triggers is achievable.

The hardware and software integration into the LCT camera of the previously described system based on a single Virtex-4FX12 FPGA 
has been achieved. 
The L2 system still needs to be fully integrated in the H.E.S.S. acquisition and tested with real data. 
This will be achieved at the beginning of the HESS-2 phase.

\section*{Acknowledgments}
The authors are grateful to D.~Besin and H.~Zaghia for the layout of the cPCI board, to P.~Nayman for providing the firmware for the PCI bridge and to the anonymous reviewers for their  valuable comments and suggestions. 

%-------------------------------
%Bibliography
%-------------------------------


\begin{thebibliography}{99}
\bibitem{MAGICCrab} E.Aliu et al, Science, 322,1221 (2008)
\bibitem{MAGICtiming} E.Aliu et al, Astropart Phys. 30, 293 (2009)
\bibitem{Bastieri2001} D.Bastieri et al, Nucl. Instrum. Meth. A461, 521 (2001)
\bibitem{Cortina2001} J. Cortina, J.C.Gonzalez, Astropart. Phys. 15, 203 (2001)
\bibitem{Funk2005}S.Funk et al, Astropart.Phys. 22, 285 (2004)
\bibitem{Guy2002} J.Guy, P.Vincent, J-P. Tavernet \& M.Rivoal, Astropart. Phys.17, 409 (2002)
\bibitem{Guythesis}J.Guy, PhD. thesis, Universit\'e Pierre et Marie Curie, (2003)
\bibitem{KASKADE} M.P. Kertzman \& G.H Sembroski, Nucl. Instrum.Meth. A343, 629 (1994)
\bibitem{Preuss2002}S.Preuss et al, Nucl.Instrum.Meth.A481,229 (2002)
\bibitem{CPCINORME} CompactPCI Specification - PICMG 2.0 R3.0  (1999)
\bibitem{hillas} A.M.Hillas, Space Science Reviews, 75, 17 (1996)
\bibitem{apu} H.H.Ng and L.Pillai, Xilinx Application Notes, 717 (2005)
\end{thebibliography}
\end{document}